%% file: substructure.tex
\documentstyle[epsfig,my,cite]{aipproc}

\def\figdir{figures/}

\input{definitions.tex}

\begin{document}

\title{Jet Substructure at HERA\footnotemark[1]}

\author{Claudia Glasman$^*$\\
representing the ZEUS Collaboration}

\address{$^*$Department of Physics and Astronomy, Kelvin Building,\\
University of Glasgow, Glasgow, G12 8QQ, UK}

\maketitle

\begin{abstract}
Measurements of jet shapes and subjet multiplicities in photoproduction
performed by ZEUS are presented and compared to leading-logarithm
parton-shower Monte Carlo models. The predicted differences on the size of
gluon- and quark-initiated jets are used to select samples to study the
dynamics of the subprocesses.
\end{abstract}

\fntext{}{\talk{International Conference on the Structure and Interactions of
the Photon {\rm (PHOTON 2000)}}{Ambleside}{UK}{August $26^{th}-31^{st}$}{2000}}

\section*{Introduction}

The internal structure of a jet depends mainly on the type of primary parton
(quark or gluon) from which it originated and to a lesser extent on the
particular hard scattering process. QCD predicts that (a) at sufficiently high
transverse energy of the jet ($\etjet$), where fragmentation effects become
negligible, the jet structure is driven by gluon emission from the primary
parton; and (b) gluon jets are broader than quark jets due to the larger
colour charge of the gluon. In the first part of this article, the
measurements performed by ZEUS \cite{shape} to test the QCD prediction (a)
are presented and compared to leading-logarithm parton-shower Monte Carlo
models. The lowest non-trivial order contribution to the jet substructure is
given by ${\cal O}(\alpha\alpha_s^2)$ calculations and, therefore measurements
of jet substructure provide a stringest test of QCD beyond leading order (LO).
The results of taking prediction (b) into account to test the dynamics of the
subprocesses \cite{substr} are shown in the second part of this article.

\begin{figure*}
\begin{center}
\setlength{\unitlength}{1.0cm}
\begin{picture} (5.0,4.0)
\put (-3.0,0.2){\epsfig{figure=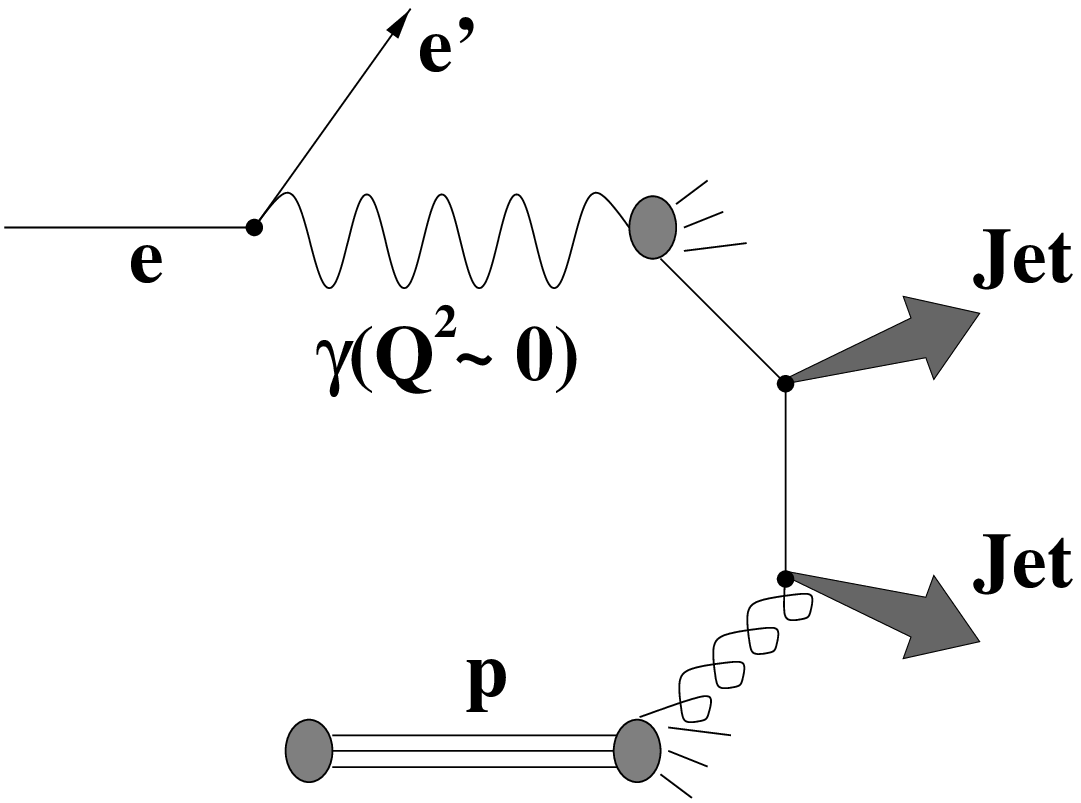,width=5cm}}
\put (4.0,0.2){\epsfig{figure=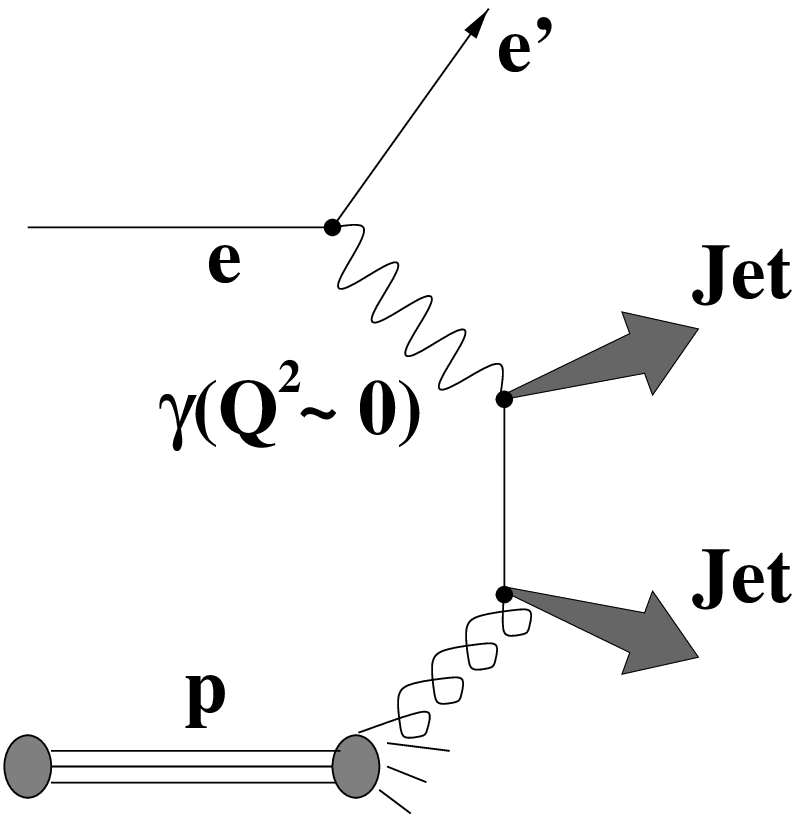,width=4cm}}
\put (-0.2,-0.3){\small (a)}
\put (5.5,-0.3){\small (b)}
\end{picture}
\end{center}
\caption{\label{fig1} (a) Resolved- and (b) direct-photon processes.}
\end{figure*}

At HERA, positrons of energy $E_e=27.5$~GeV collide with protons of energy
$E_p=820$~GeV. The main source of jets is hard scattering in $\gp$
interactions in which a quasi-real photon ($\q2\approx 0$, where $\q2$ is the
virtuality of the photon) emitted by the positron beam interacts with a parton
from the proton to produce two jets in the final state. In LO QCD, there are
two processes which contribute to the jet production cross section: the
resolved process (figure~\ref{fig1}a) in which the photon interacts through
its partonic content, and the direct process (figure~\ref{fig1}b) in which
the photon interacts as a point-like particle.

%Figure 1

In the kinematic regime studied here, the dominant subprocesses are
$q_{\gamma}g_p\rightarrow qg$ in resolved and $\gamma g\rightarrow q\bar q$ in
direct. The kinematics of these processes are such that the majority of the
jets in the region of pseudorapitity$^{\ddag}$\footnote{The pseudorapidity is
defined as $\eta=-\ln(\tan\frac{\theta}{2})$, where the polar angle $\theta$
is taken with respect to the proton beam direction. The ZEUS coordinate system
is defined as right-handed with the $Z$-axis pointing in the proton beam
direction, hereafter referred to as forward, and the $X$-axis horizontal,
pointing towards the centre of HERA.} ($\etajet$) below 0 originate from
quarks and the fraction of gluon-initiated jets increases as $\etajet$
increases.

\section*{Jet substructure}

The internal structure of jets can be studied by means of the jet shape and
the subjet multiplicity.

\subsection*{Jet shape}

The differential jet shape is defined as the average fraction of the jet's
transverse energy that lies inside an annulus in the pseudorapidity ($\eta$) -
azimuth ($\varphi$) plane of radii
$r\pm\Delta r/2$, where $r=\sqrt{(\Delta\eta)^2+(\Delta\varphi)^2}$,
concentric with the jet axis and is given by

$$\ro={1\over N_{jets}}\cdot{1\over\Delta r}\cdot
\displaystyle\sum_{jets}
{E_T(r-{\Delta r\over 2},r+{\Delta r\over 2})\over E_T(0,r=1)},$$
where $E_T(r-\Delta r/2,r+\Delta r/2)$ is the transverse energy within the 
given annulus and $N_{jets}$ is the total number of jets in the sample.
The integrated jet shape is the average fraction of the jet's transverse
energy that lies inside a cone in the $\etaphi$ plane of radius $r$ concentric
with the jet axis,

$$\ps={1\over N_{jets}}\cdot\displaystyle\sum_{jets}{E_T(r)\over 
E_T(r=1)}.$$

\begin{figure*}
\begin{center}
\setlength{\unitlength}{1.0cm}
\begin{picture} (10.0,10.0)
\put (-3.8,-1.0){\epsfig{figure=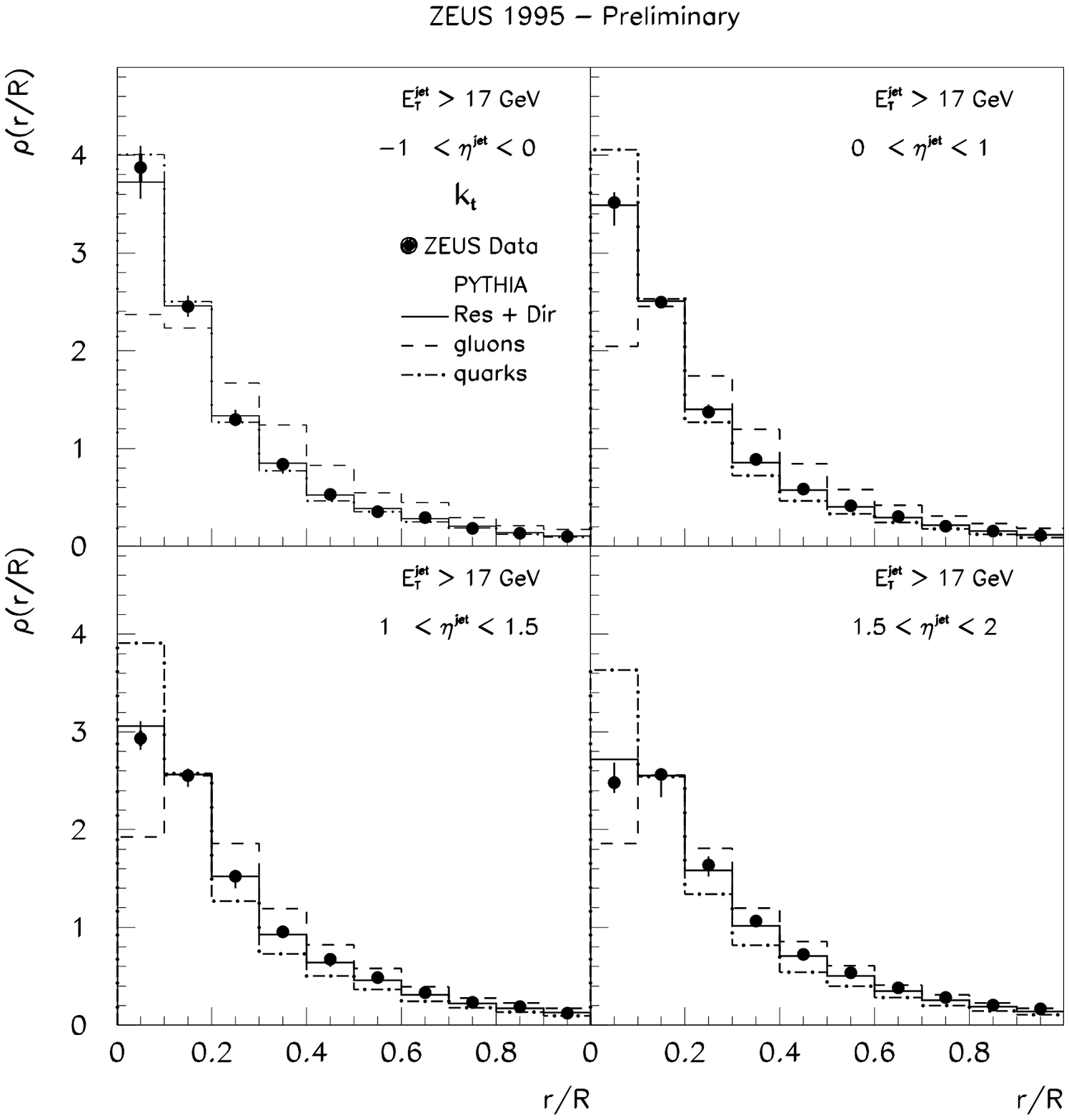,width=10cm}}
\put (-3.5,5.5){\epsfig{figure=\figdir white.eps,width=0.5cm,height=5cm}}
\put (1.0,1.5){\epsfig{figure=\figdir white.eps,width=5cm,height=0.5cm}}
\put (0.4,9.1){\epsfig{figure=\figdir white.eps,width=0.4cm,height=0.4cm}}
\put (-3.44,1.53){\epsfig{figure=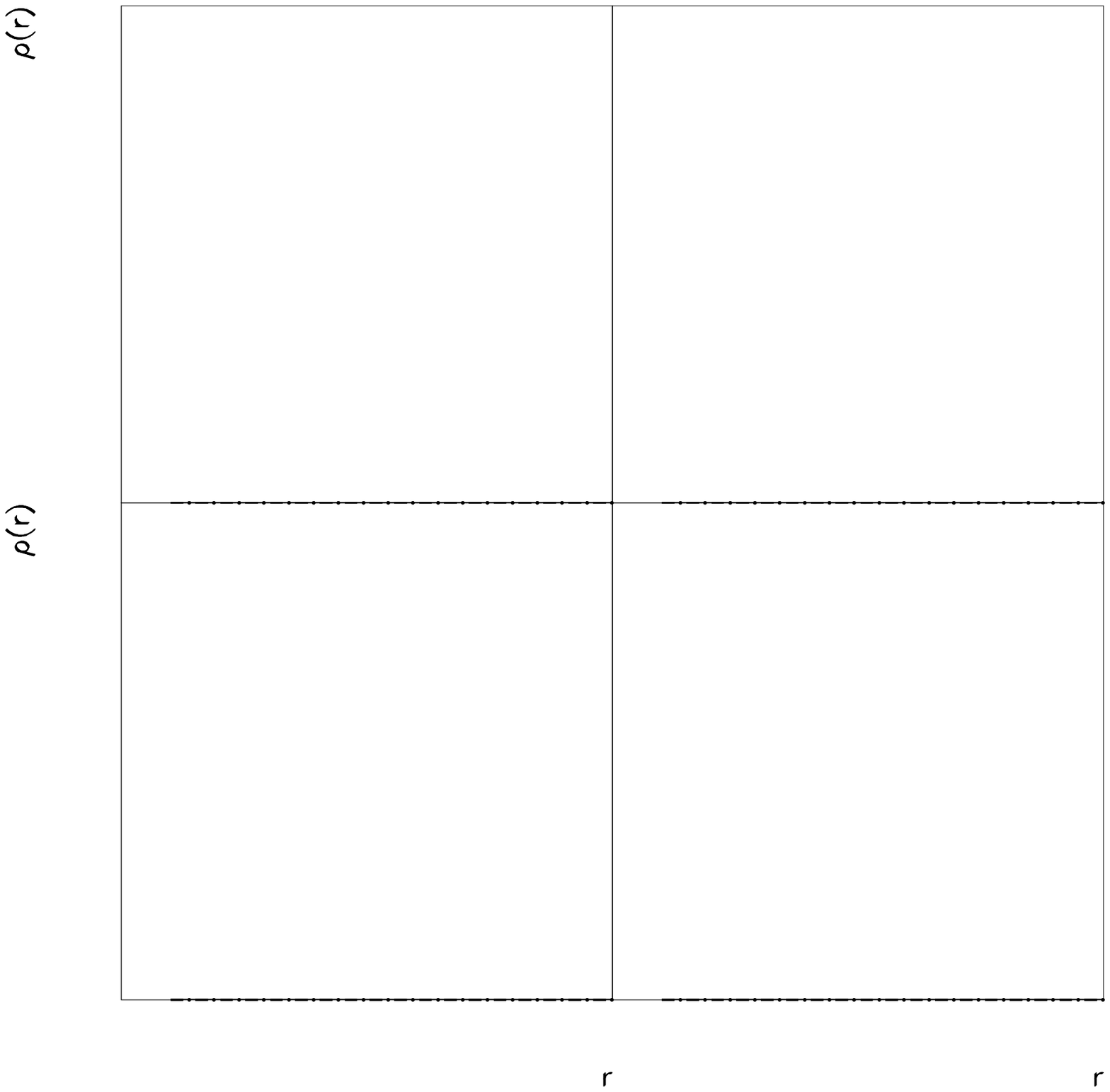,width=9.8cm}}
\put (5.5,2.5){\epsfig{figure=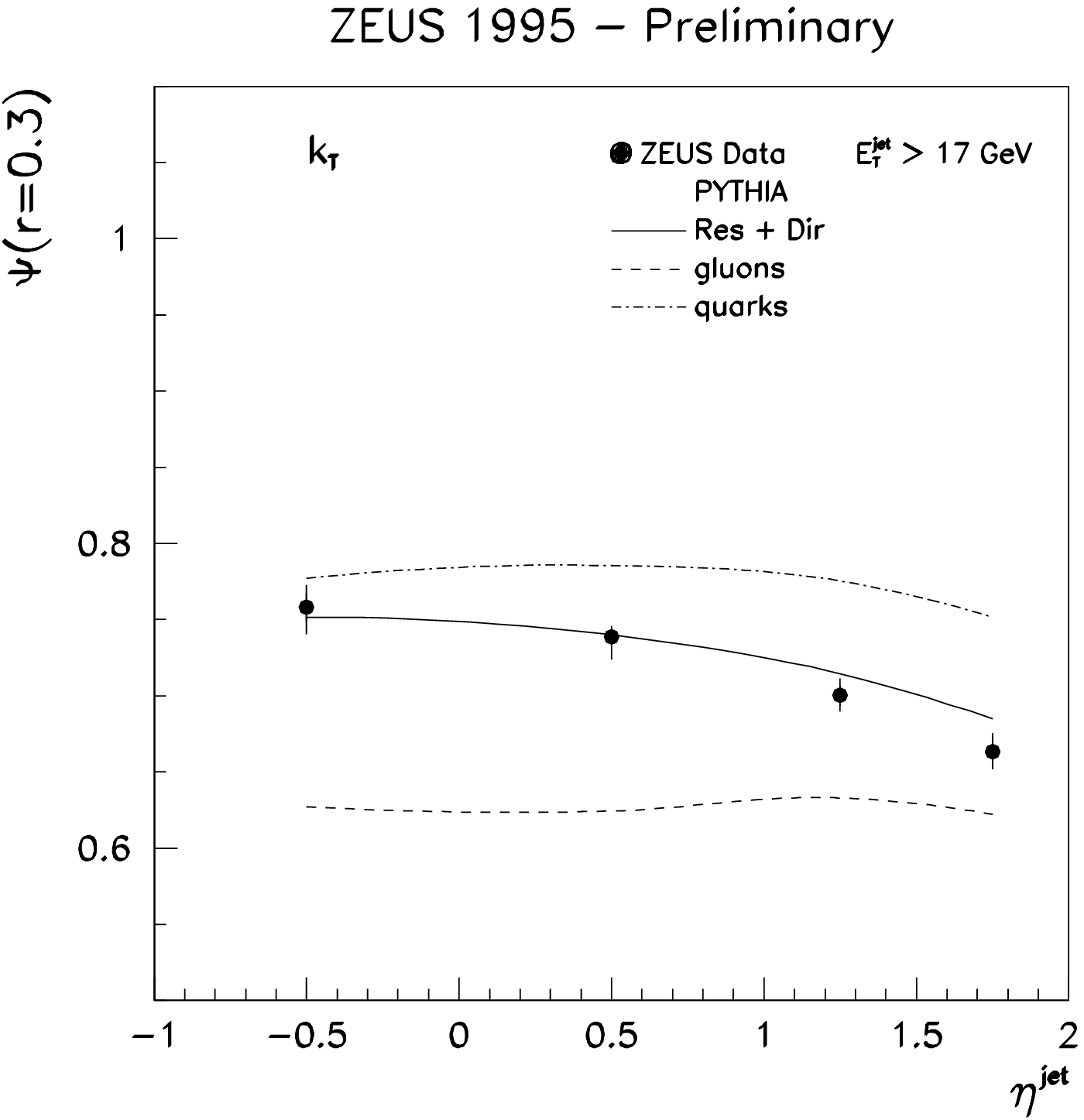,width=8cm}}
\put (7.7,9.4){\epsfig{figure=\figdir white.eps,width=0.4cm,height=0.4cm}}
\put (1.2,1.4){\small (a)}
\put (9.0,1.4){\small (b)}
\end{picture}
\end{center}
\vspace{-1.5cm}
\caption{\label{fig2} (a) Measured differential jet shape $\ro$ vs. $r$
(black dots). (b) Measured integrated jet shape $\psi(r=0.3)$ vs. $\etajet$
(black dots). PYTHIA Monte Carlo calculations are shown for comparison.}
\end{figure*}

Measurements of the differential jet shape $\ro$ have been performed by
ZEUS~\cite{shape} using an inclusive sample of jets. The jets have been
identified using the longitudinally invariant $\kt$ cluster algorithm
\cite{kt} in the inclusive mode. The jets were required to have
$\etjet>17$~GeV and $-1<\etajet<2$. The measurements were performed in the
kinematic region $0.2<y<0.85$, where $y$ is the inelasticity variable, and
$\q2\leq 1$~\g2. Figure~\ref{fig2}a shows $\ro$ as a function of $r$ in
different regions of $\etajet$. The data show that the jets become broader as
$\etajet$ increases.

%Figure 2

The solid histograms in figure~\ref{fig2}a are calculations from the
leading-logarithm parton-shower Monte Carlo PYTHIA~\cite{pythia}. The
predictions, which include initial and final state QCD radiation, give a good
description of the data. The comparison of the predictions for gluon- (dashed
histograms) and quark-initiated (dot-dashed histograms) jets to the data shows
that the measured jets are quark-like for $-1<\etajet<0$ and become
increasingly more gluon-like as $\etajet$ increases.

The quark and gluon content of the final state has been investigated in
more detail by studying the $\etajet$ dependence of the integrated jet shape
at a fixed value of $r$ \cite{shape}. Figure~\ref{fig2}b shows $\ps$ for
$r=0.3$ as a function of $\etajet$. The measured jet shape decreases with
$\etajet$, i.e. the jets become broader as $\etajet$ increases. The comparison
between the predictions of PYTHIA for gluon- and quark-initiated jets and the
data shows that the broadening of the jets is consistent with an increasing
fraction of gluon-initiated jets as $\etajet$ increases.

\subsection*{Subjet multiplicity}

The application of the $\kt$ cluster algorithm allows the study of the internal
structure of the jets by means of jet-like objects (subjets) for which resummed
QCD calculations are possible. Subjets are resolved within a jet by reapplying
the $\kt$ cluster algorithm until for every pair of particles the quantity 
$$d_{ij}=\min(E_{Ti},E_{Tj})^2\cdot[(\eta_i-\eta_j)^2+(\varphi_i-\varphi_j)^2]$$
is above $\yc\cdot(\etjet)^2$. The theoretical advantages of this
observable are that safe observables can be defined at any order in
perturbative QCD and resummed calculations are possible. Subjets are also a
useful tool to investigate the colour dynamics. Finally, the uncertainties
coming from hadronisation corrections are small when $\yc$ is not too small.

\begin{figure*}
\begin{center}
\setlength{\unitlength}{1.0cm}
\begin{picture} (10.0,8.0)
\put (-4.0,0.0){\epsfig{figure=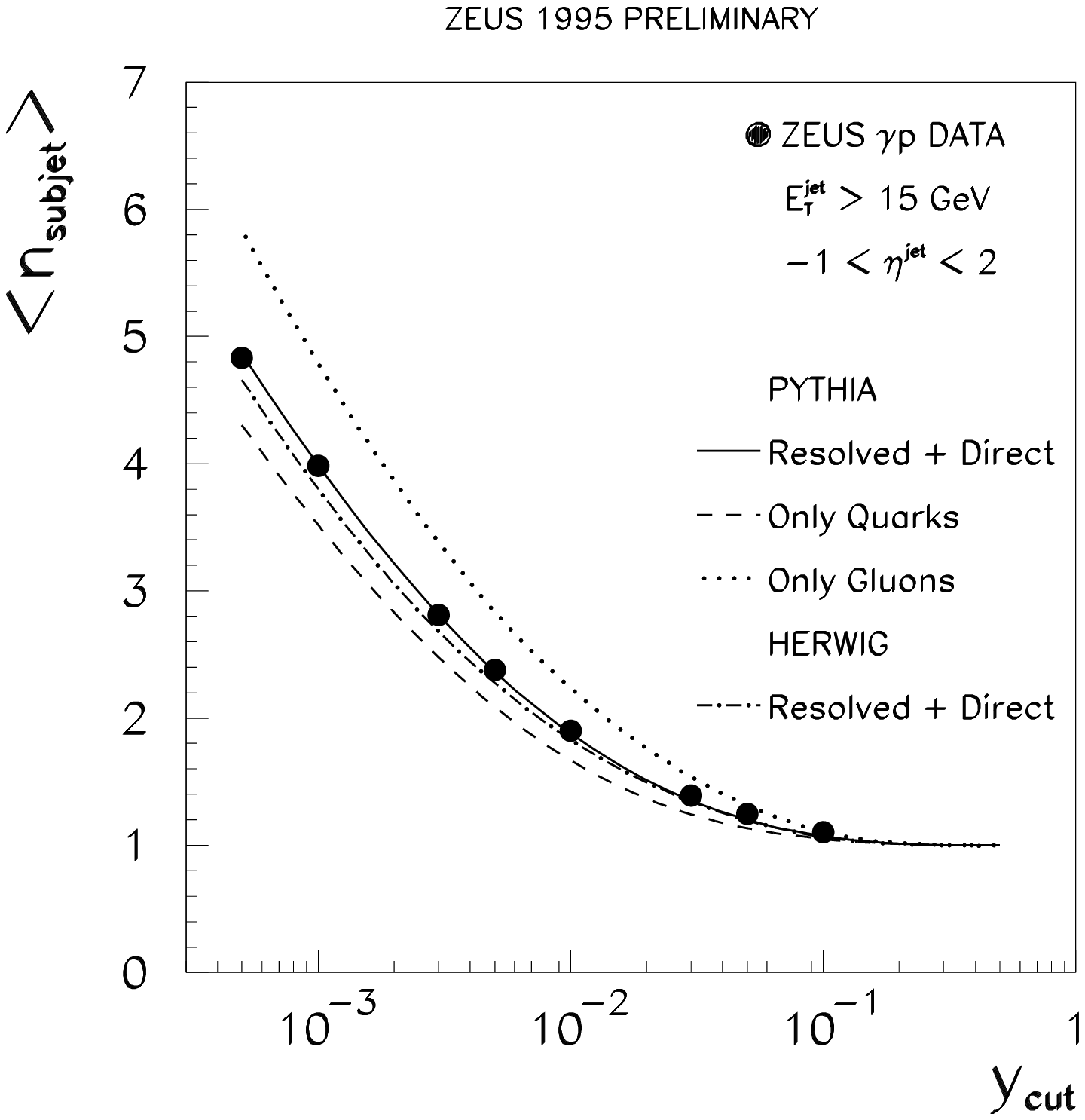,width=10cm}}
\put (4.0,0.0){\epsfig{figure=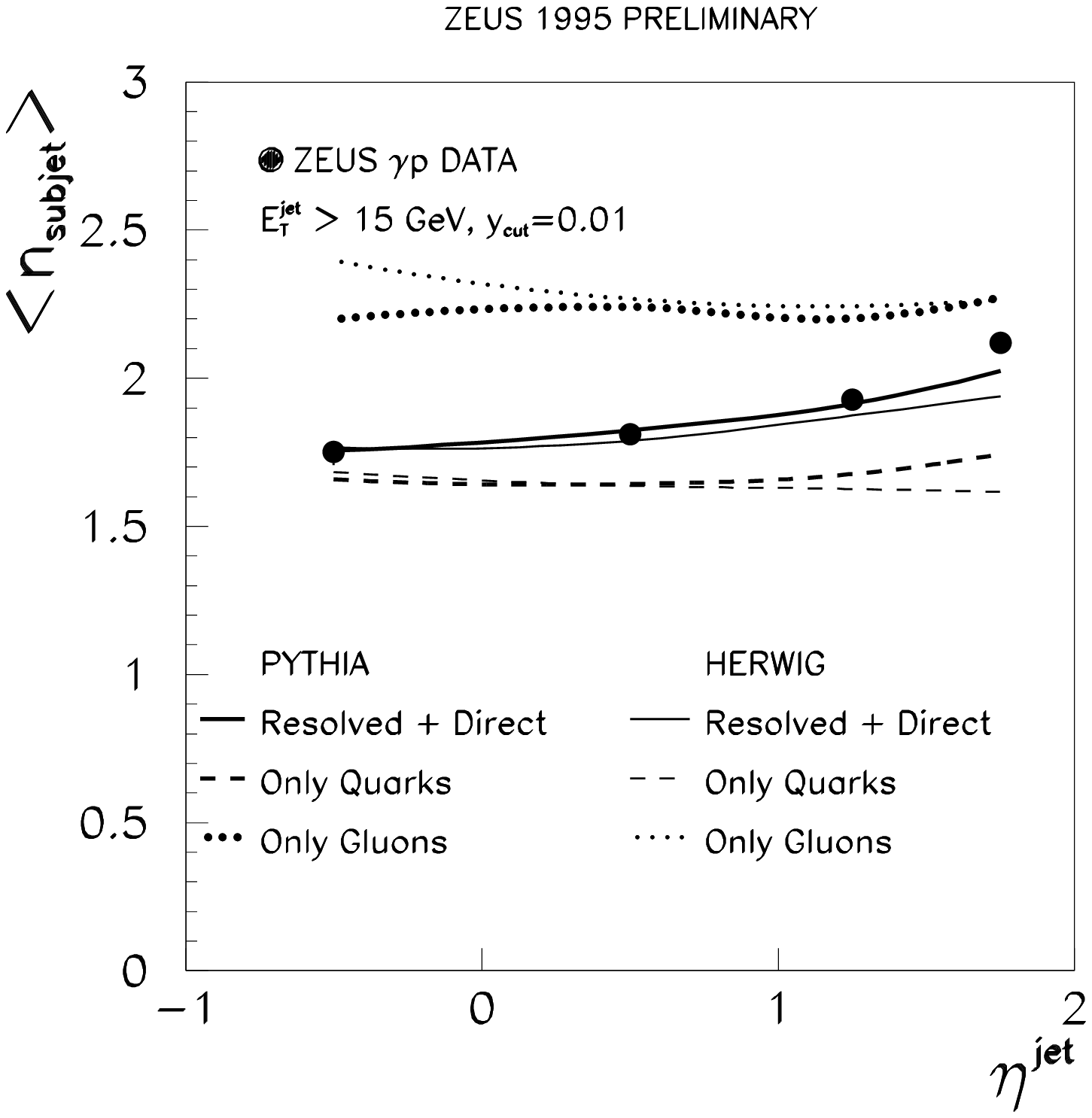,width=10cm}}
\put (-1.5,7.5){\small (a)}
\put (11.0,7.5){\small (b)}
\end{picture}
\end{center}
\vspace{-1.0cm}
\caption{\label{fig3} (a) Measured mean subjet multiplicity $<\ns>$ vs.
$\yc$ (black dots). (b) Measured mean subjet multiplicity $<\ns(\yc=0.01)>$ vs.
$\etajet$ (black dots). PYTHIA and HERWIG Monte Carlo calculations are shown
for comparison.}
\end{figure*}

The mean subjet multiplicity $<\ns>$ has been measured \cite{shape} for an
inclusive sample of jets with $\etjet>15$ GeV and $-1<\etajet<2$ in the
kinematic region given by $0.2<y<0.85$ and $\q2\leq 1$~\g2. Figure~\ref{fig3}a
shows $<\ns>$ as a function of $\yc$. The predictions of PYTHIA and
HERWIG~\cite{herwig}, which include initial and final state QCD radiation,
provide a good description of the data. The measured  $<\ns>$ for a fixed
$\yc$ value of 0.01 (see figure~\ref{fig3}b) increases as $\etajet$
increases. The comparison to the predictions for gluon- and quark-initiated
jets shows that the increase in  $<\ns>$ as $\etajet$ increases is consistent
with the predicted increase in the fraction of gluon-initiated jets. This
result is consistent with the one obtained from the measurements of jet shapes.

%Figure 3

\section*{Substructure dependence of dijet cross sections}

The predictions of the Monte Carlo models for the jet shape and subjet
multiplicity reproduce the data well and show the expected differences for
quark- and gluon-initiated jets. Therefore, the Monte Carlo events have been
used to devise a method to select samples enriched in quark- and
gluon-initiated jets. The samples are selected by exploiting the QCD
prediction that gluon-initiated jets should be broader than quark-initiated
jets.

\begin{figure*}
\begin{center}
\setlength{\unitlength}{1.0cm}
\begin{picture} (10.0,8.0)
\put (-4.0,0.0){\epsfig{figure=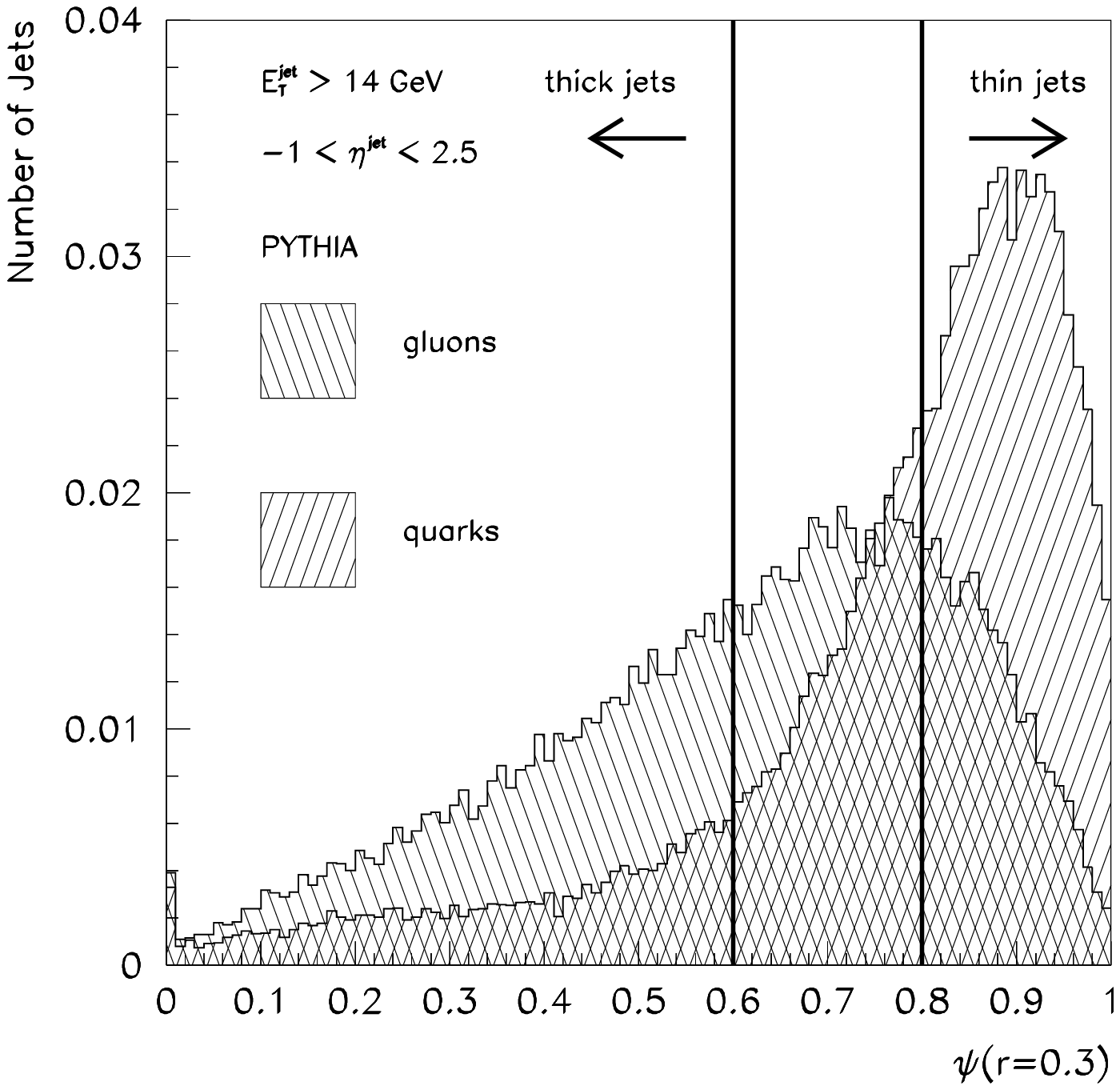,width=10cm}}
\put (5.25,1.25){\epsfig{figure=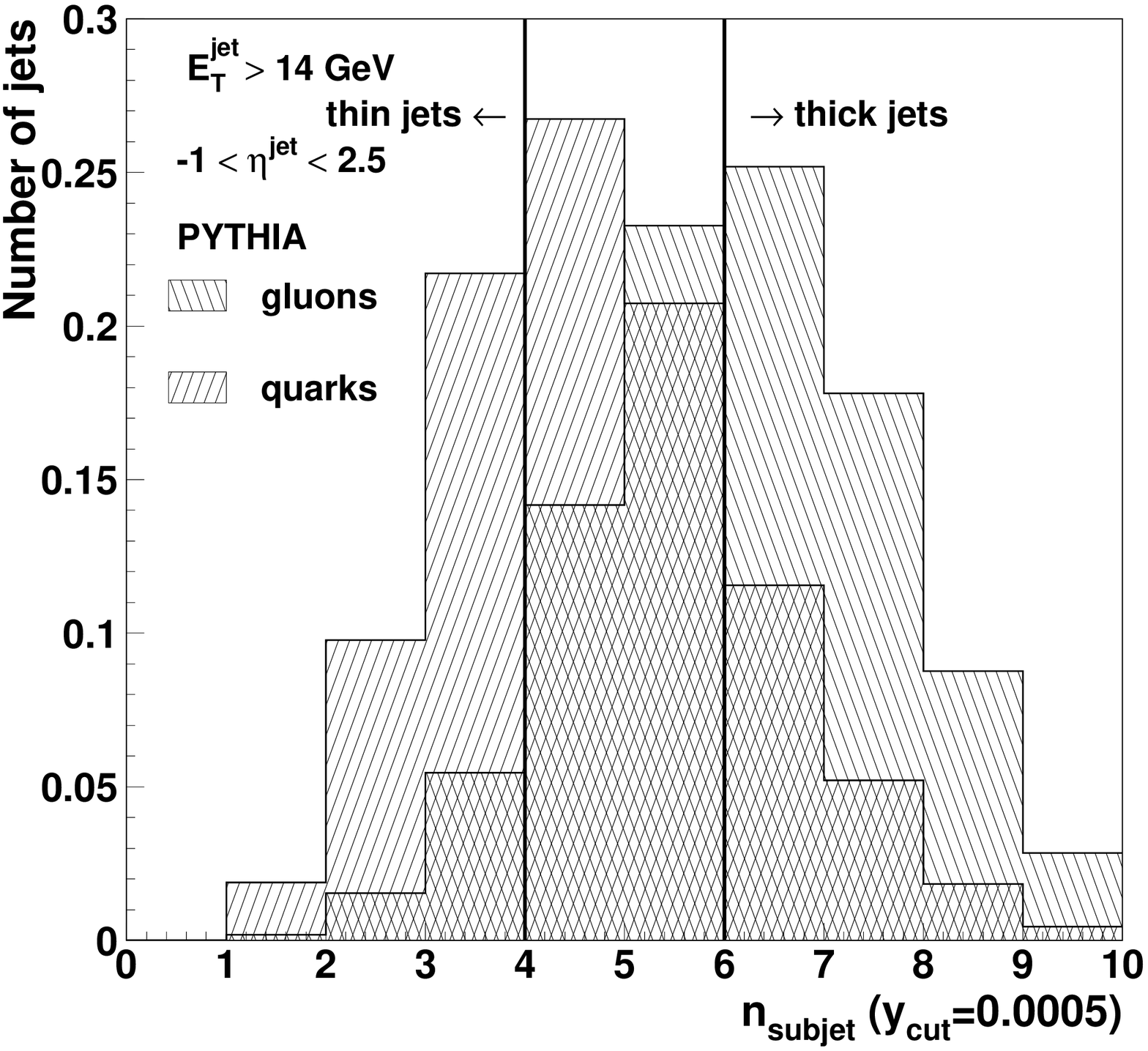,width=6.92cm,height=7.5cm}}
\put (5.1,1.0){\epsfig{figure=\figdir white.eps,width=1.1cm,height=7.5cm}}
\put (5.5,0.85){\epsfig{figure=\figdir white.eps,width=7cm,height=1.1cm}}
\put (5.8,5.0){\epsfig{figure=\figdir white.eps,width=1.9cm,height=3.5cm}}
\put (7.3,7.2){\epsfig{figure=\figdir white.eps,width=1cm,height=0.5cm}}
\put (9.7,7.2){\epsfig{figure=\figdir white.eps,width=1.6cm,height=0.5cm}}
\put (4.0,0.0){\epsfig{figure=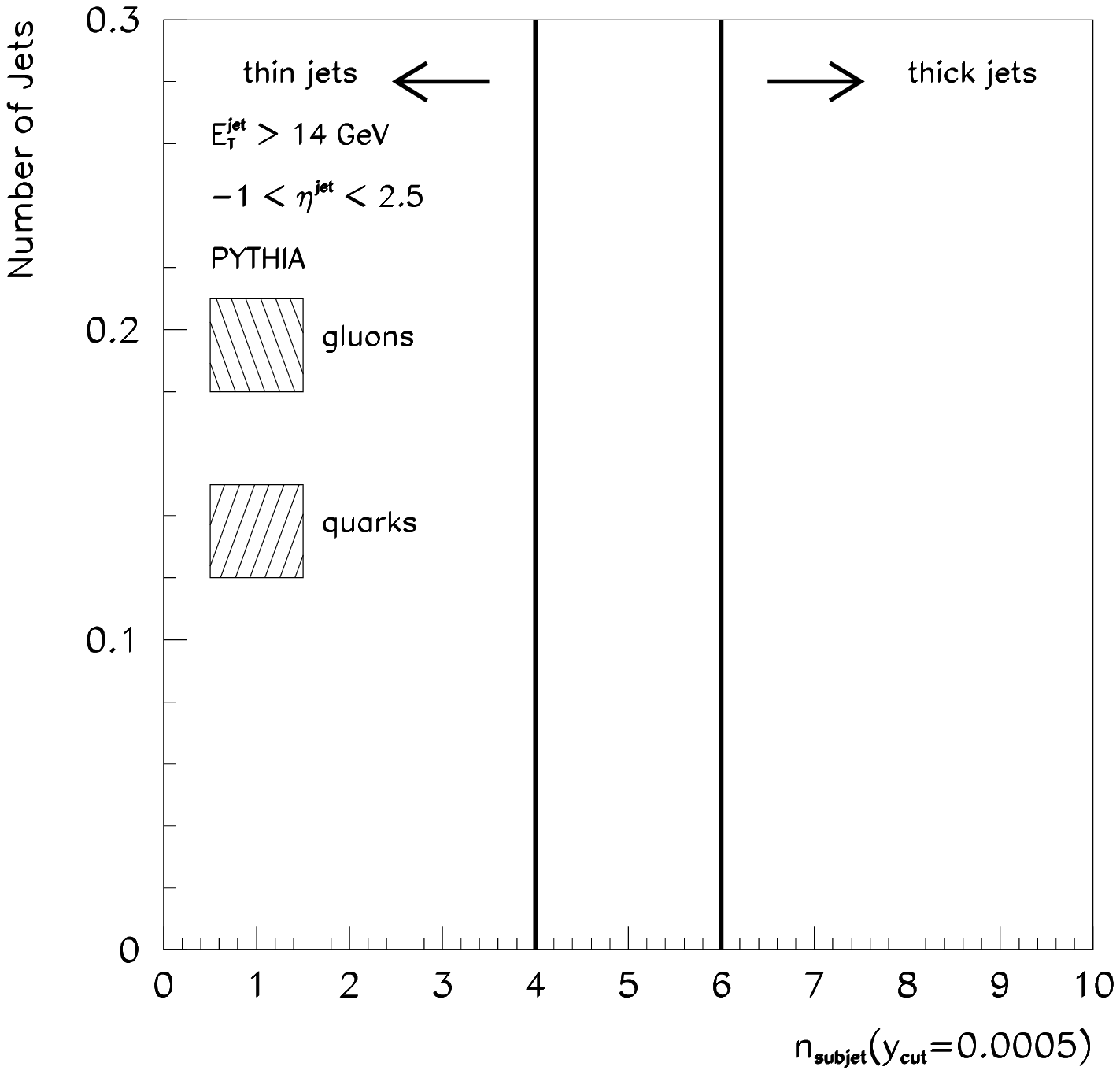,width=10cm}}
\put (-1.9,7.0){\small (a)}
\put (11.0,7.0){\small (b)}
\end{picture}
\end{center}
\vspace{-1.0cm}
\caption{\label{fig4} The predicted integrated jet shape $\psi(r=0.3)$ (a)
and subjet multiplicity $\ns(\yc=0.0005)$ (b) distributions at the
hadron level for samples of gluon- and quark-initiated jets simulated using the
program PYTHIA.}
\end{figure*}

The predicted distributions of $\ps$ for $r=0.3$ and $<\ns>$ for $\yc=0.0005$
for quark- and gluon-initiated jets are clearly different (see
figure~\ref{fig4}). A sample enriched in quark-initiated (``thin'') jets has
been selected by requiring $\psi(r=0.3)>0.8$ and a sample enriched in
gluon-initiated (``thick'') jets has been selected by requiring
$\psi(r=0.3)<0.6$. Alternative samples to support the results obtained with
the jet shape selection have been selected using the subjet multiplicity. Jets
are called ``thin'' if $\ns(\yc=0.0005)<4$ and the jets are called ``thick''
if $\ns(\yc=0.0005)\geq 6$.

%Figure 4

\subsection*{Measurements of $\seta$}

\begin{figure*}
\begin{center}
\setlength{\unitlength}{1.0cm}
\begin{picture} (10.0,8.0)
\put (-4.0,0.0){\epsfig{figure=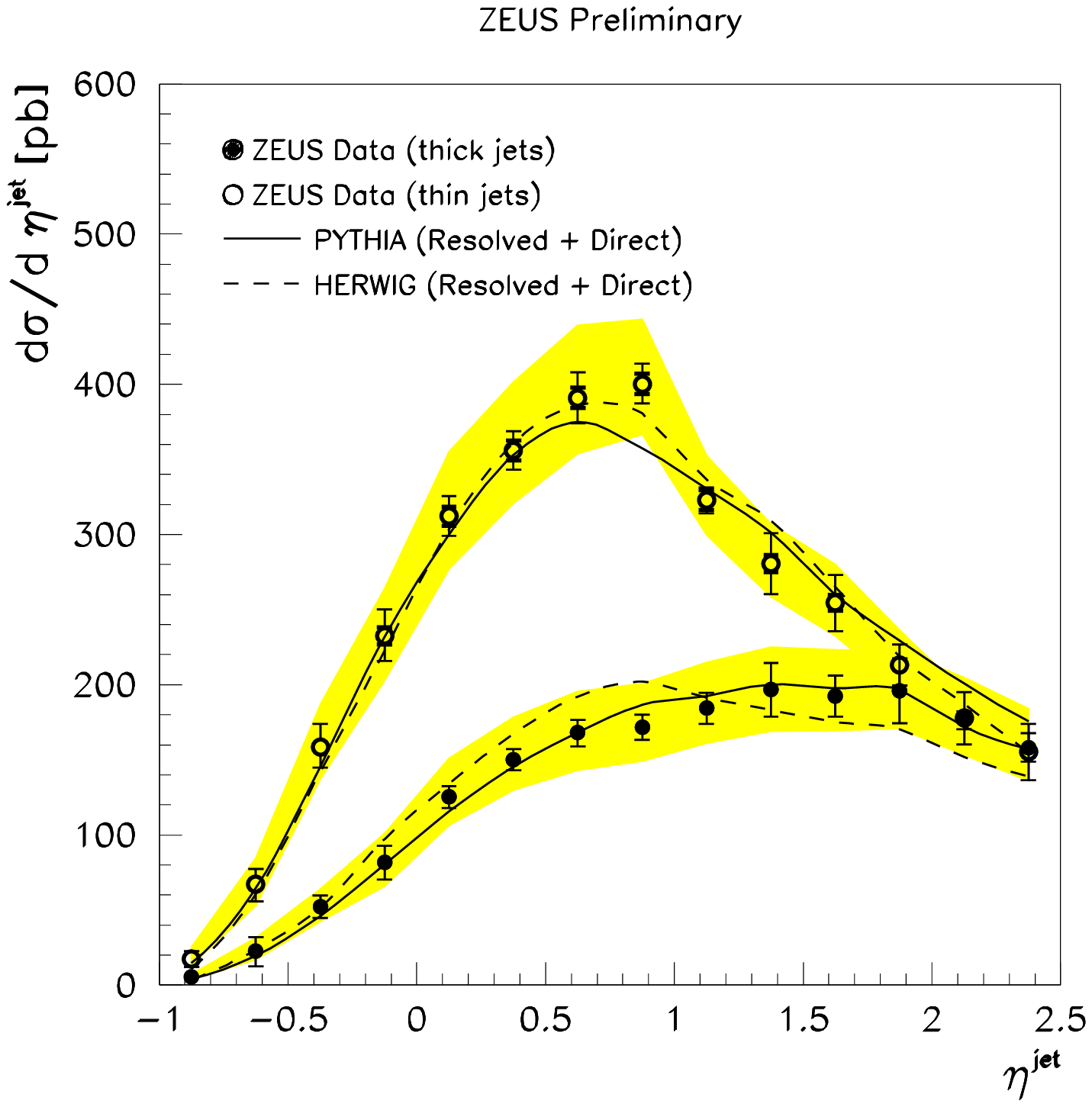,width=10cm}}
\put (4.0,0.0){\epsfig{figure=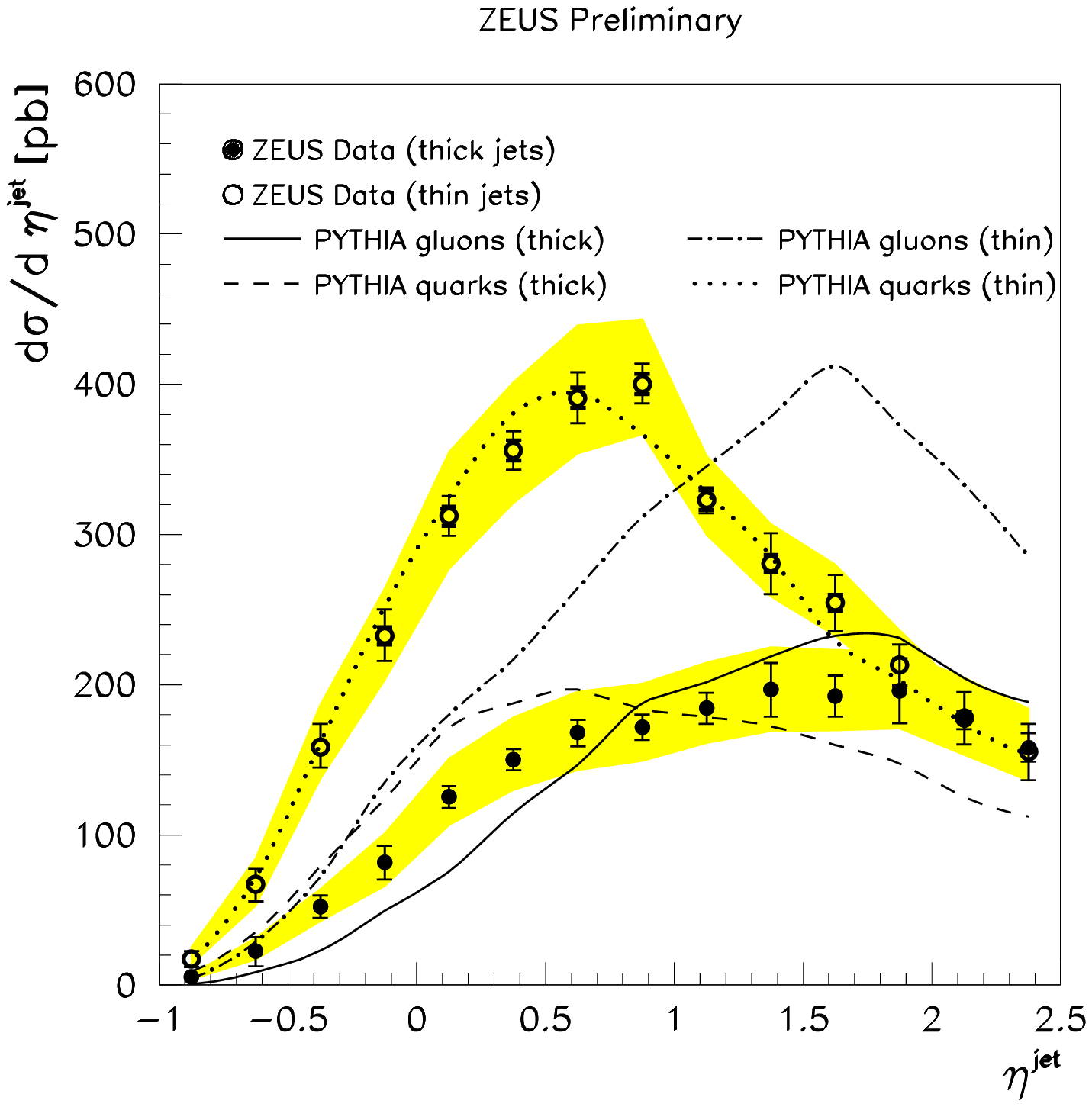,width=10cm}}
\put (3.0,7.5){\small (a)}
\put (11.0,7.5){\small (b)}
\end{picture}
\end{center}
\vspace{-1.0cm}
\caption{\label{fig5} Measured $\seta$ for samples of ``thick'' (black dots)
and ``thin'' (open circles) jets selected according to their shape. The thick
error bars represent the statistical errors of the data, and the thin error
bars show the statistical errors and uncorrelated systematic uncertainties
added in quadrature. The shaded band displays the uncertainty due to the
absolute energy scale of the jets. In (a) Monte Carlo calculations using
PYTHIA and HERWIG for resolved- plus direct-photon processes and in (b) the
predictions of PYTHIA for quark- and gluon-initiated jets are shown for
comparison.}
\end{figure*}

Using the jet shape selection into ``thin'' and ``thick'' jet samples, dijet
cross sections have been measured \cite{substr} as a function of $\etajet$ for
events with at least two jets of $\etjet>14$ GeV and $-1<\etajet<2.5$ in the
kinematic region given by $0.2<y<0.85$ and $\q2\leq 1$~\g2. Figure~\ref{fig5}
shows $\seta$ for ``thick'' and ``thin'' jets.  The cross section for ``thick''
jets displays a very different shape than that of the ``thin'' jet sample: the
$\etajet$ distribution for the ``thin'' jet sample peaks at about 0.7, whereas
the ``thick'' jet sample distribution peaks at $\etajet\approx 1.5$.

The predictions of PYTHIA and HERWIG (see figure~\ref{fig5}a) give a good
description of the data. The Monte Carlo distributions have been normalised to
the total measured cross section of each type after applying the same jet shape
selection as in the data. PYTHIA and HERWIG predict a similar parton content
of the final state: the ``thick'' jet sample is composed of $15-17\%$ of $gg$
subprocesses in the final state, $54-58\%$ of $gq$ and $25-31\%$ of $qq$. The
``thin'' jet sample contains $54-56\%$ of $qq$, $41\%$ of $qg$ and $3-5\%$ of
$gg$. Therefore, the ``thin'' jet sample is indeed dominated by
quark-initiated jets in the final state and the ``thick'' jet sample has a
high content of gluon-initiated jets coming mainly from the final-state gluon
of the subprocess $q_{\gamma}g_p\rightarrow qg$. The measurements are compared
to the predictions of quark- and gluon-initiated jets with the same jet shape
selection as for the data in figure~\ref{fig5}b. A ``thin''-quark jet sample
gives a good description of the ``thin'' jet sample, whereas the
``thin''-gluon jet sample peaks one unit of pseudorapidity higher. A sample of
``thick''-quark jets alone cannot describe the ``thick''-jet sample and a large
content of gluon-initiated jets is needed to reproduce the shape of the data.

%Figure 5

\subsection*{Measurements of $\sccos$}

The underlying parton dynamics is reflected in the distribution of the
scattering angle in the dijet centre-of-mass system, $\theta^*$. The $\ccos$
distribution is sensitive to the spin of the exchanged particle: for gluon
exchange, the cross section is proportional to $(1-\cost)^{-2}$, whereas for
quark exchange the cross section is proportional to $(1-\cost)^{-1}$. In a
previous publication \cite{angular}, the $\cost$ distribution for resolved
processes (which are dominated by subprocesses mediated by gluon exchange) was
observed to rise more steeply than that of direct processes (which proceed via
quark exchange) at high $\cost$ values. In this analysis \cite{substr},
samples of jets are selected according to their internal structure. This
method provides a handle to better understand the dynamics of the subprocesses.

%Figure 6
\begin{figure*}
\begin{center}
\setlength{\unitlength}{1.0cm}
\begin{picture} (10.0,8.0)
\put (-4.0,0.0){\epsfig{figure=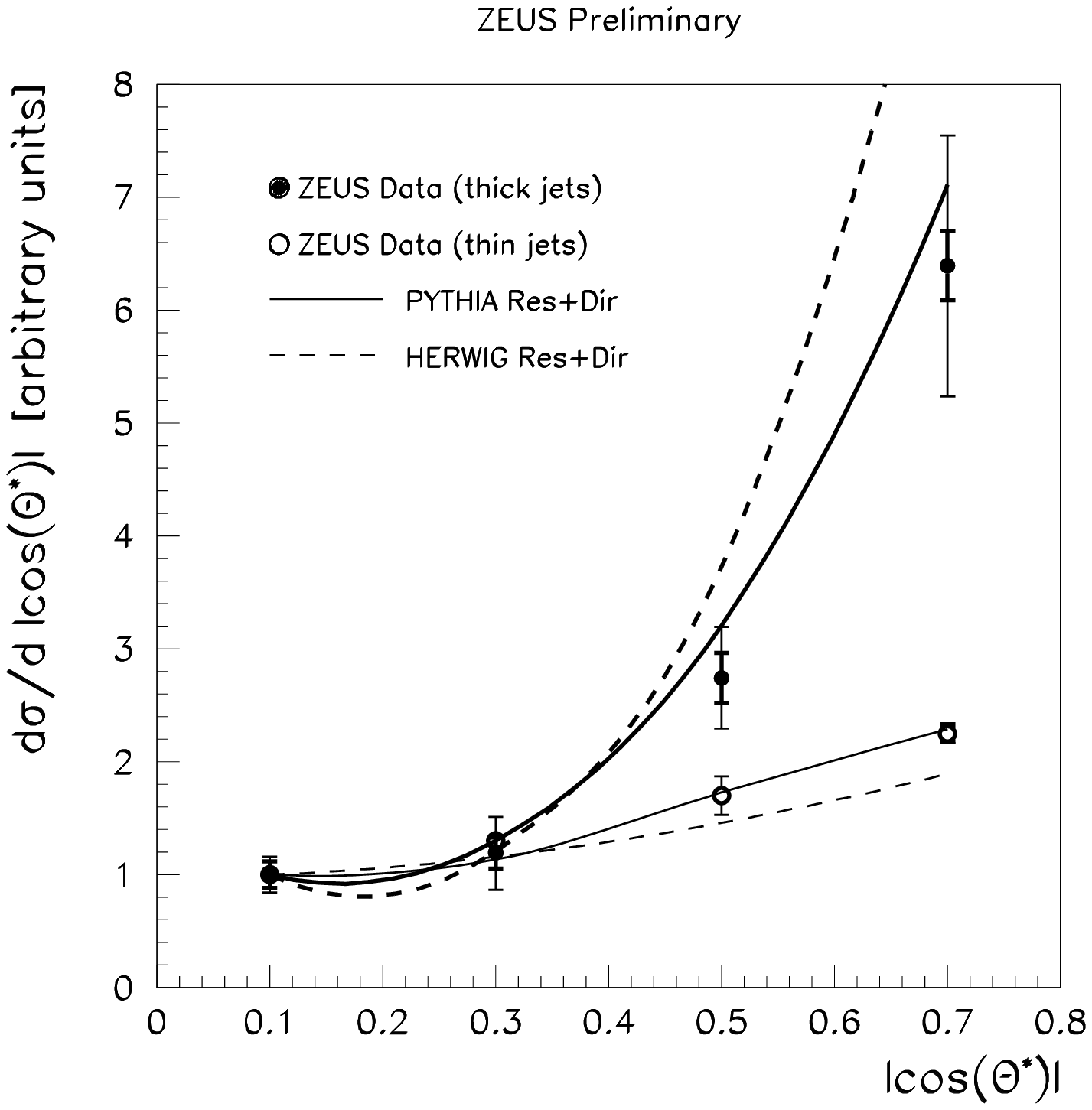,width=10cm}}
\put (5.25,1.24){\epsfig{figure=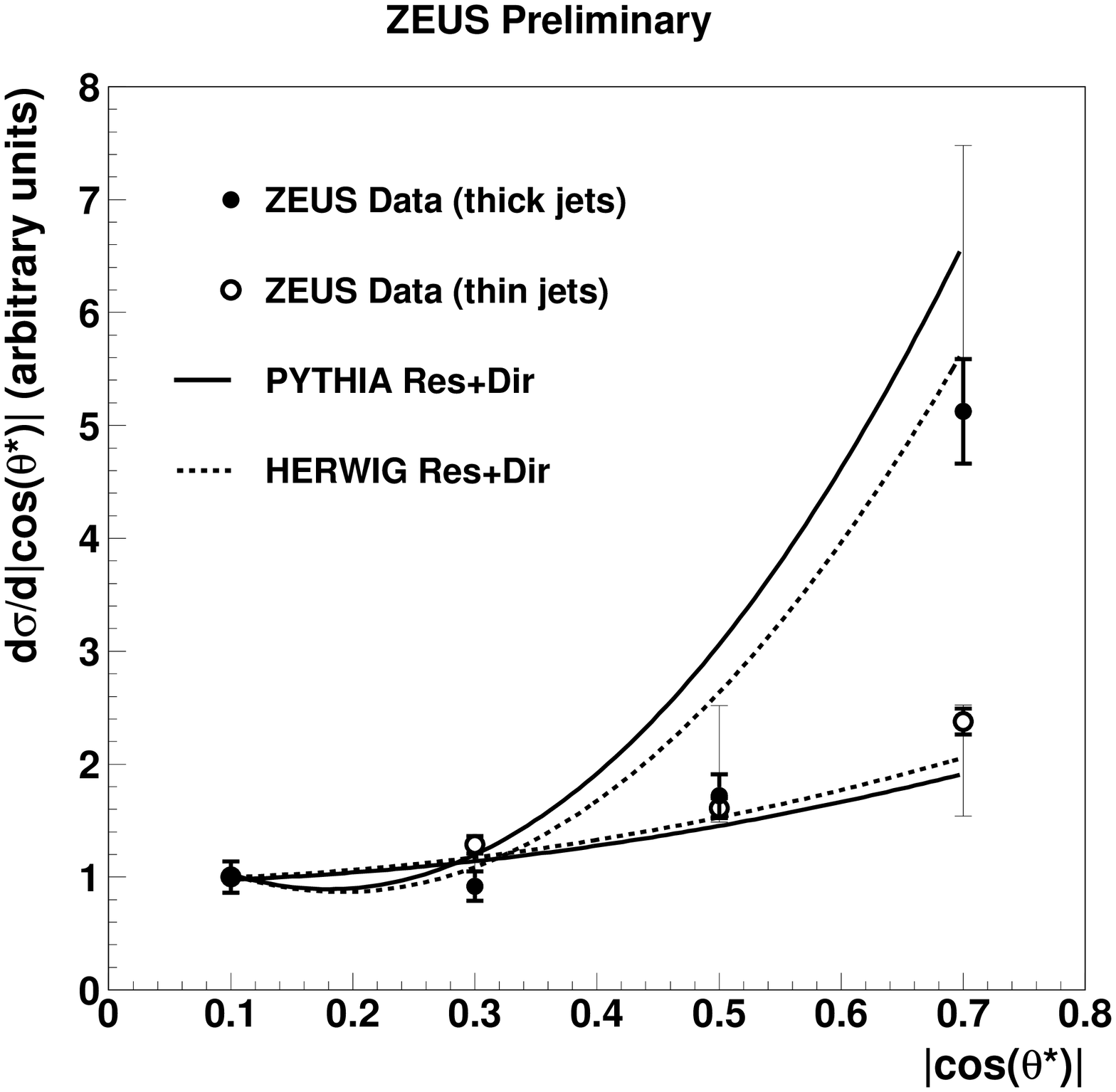,width=6.92cm,height=7.51cm}}
\put (5.1,1.0){\epsfig{figure=\figdir white.eps,width=1.1cm,height=8cm}}
\put (5.5,1.0){\epsfig{figure=\figdir white.eps,width=7cm,height=1.1cm}}
\put (6.2,4.7){\epsfig{figure=\figdir white.eps,width=3cm,height=3cm}}
\put (7.0,8.2){\epsfig{figure=\figdir white.eps,width=3.5cm,height=0.5cm}}
\put (4.0,0.0){\epsfig{figure=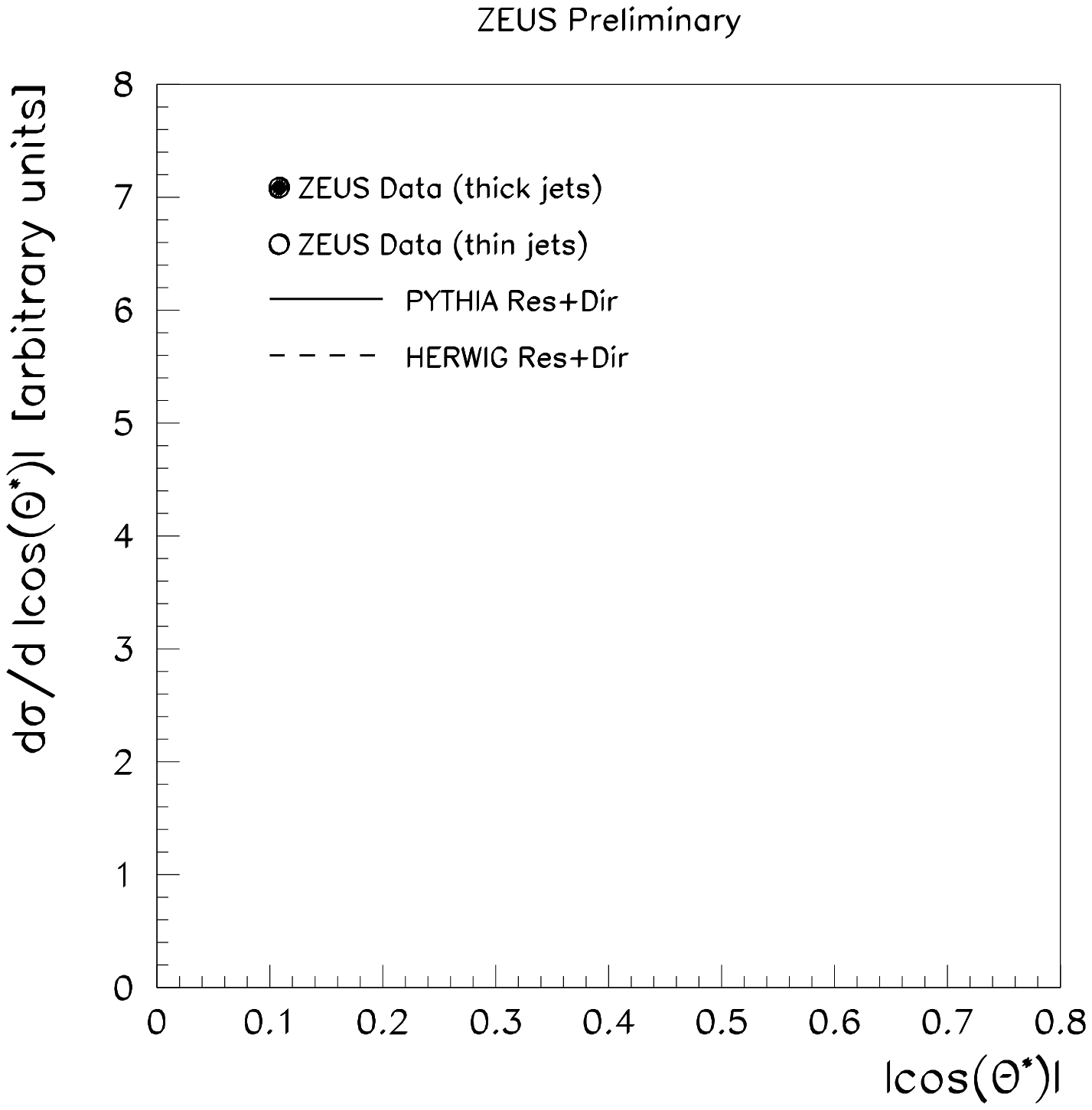,width=10cm}}
\put (-1.7,7.5){\small (a)}
\put (6.4,7.5){\small (b)}
\end{picture}
\end{center}
\vspace{-1.0cm}
\caption{\label{fig6} Measured $\scost$ for two ``thick''-jet events (black
dots) and two ``thin''-jet events (open circles) selected according to their
shape (a) or to their subjet multiplicity (b). PYTHIA and HERWIG Monte Carlo
calculations are shown for comparison.}
\end{figure*}

For samples of two ``thick'' or two ``thin'' jets, only the absolute value of
$\ccos$ can be determined because the outgoing jets are indistinguishable.
The dijet cross section as a function of $\cost$ for samples of two
``thick''-jet events and two ``thin''-jet events has been measured for dijet
invariant masses $\mj>47$ GeV. The cross sections are presented in
figure~\ref{fig6} and are normalised so as to have a value of unity at low
$\cost$ to study the difference in slope as $\cost\rightarrow 1$. The $\cost$
distribution for the two samples of dijet events increases as $\cost$
increases, however they exhibit a different slope. For comparison, PYTHIA and
HERWIG Monte Carlo predictions are shown in figure~\ref{fig6}. The
predictions have been also normalised so as to have a value of unity at low
$\cost$ and give a reasonable description of the shape of the data. The
different slope observed in the data can be understood in terms of the
dominant subprocesses in the two samples: the two ``thick''-jet sample is
dominated by subprocesses mediated by gluon exchange ($gg\rightarrow gg$ and
$qg\rightarrow qg$) whereas the two ``thin''-jet sample is dominated by
subprocesses mediated by quark exchange ($\gamma g\rightarrow q\bar q$).

\begin{figure*}
\begin{center}
\setlength{\unitlength}{1.0cm}
\begin{picture} (10.0,8.0)
\put (-4.0,0.0){\epsfig{figure=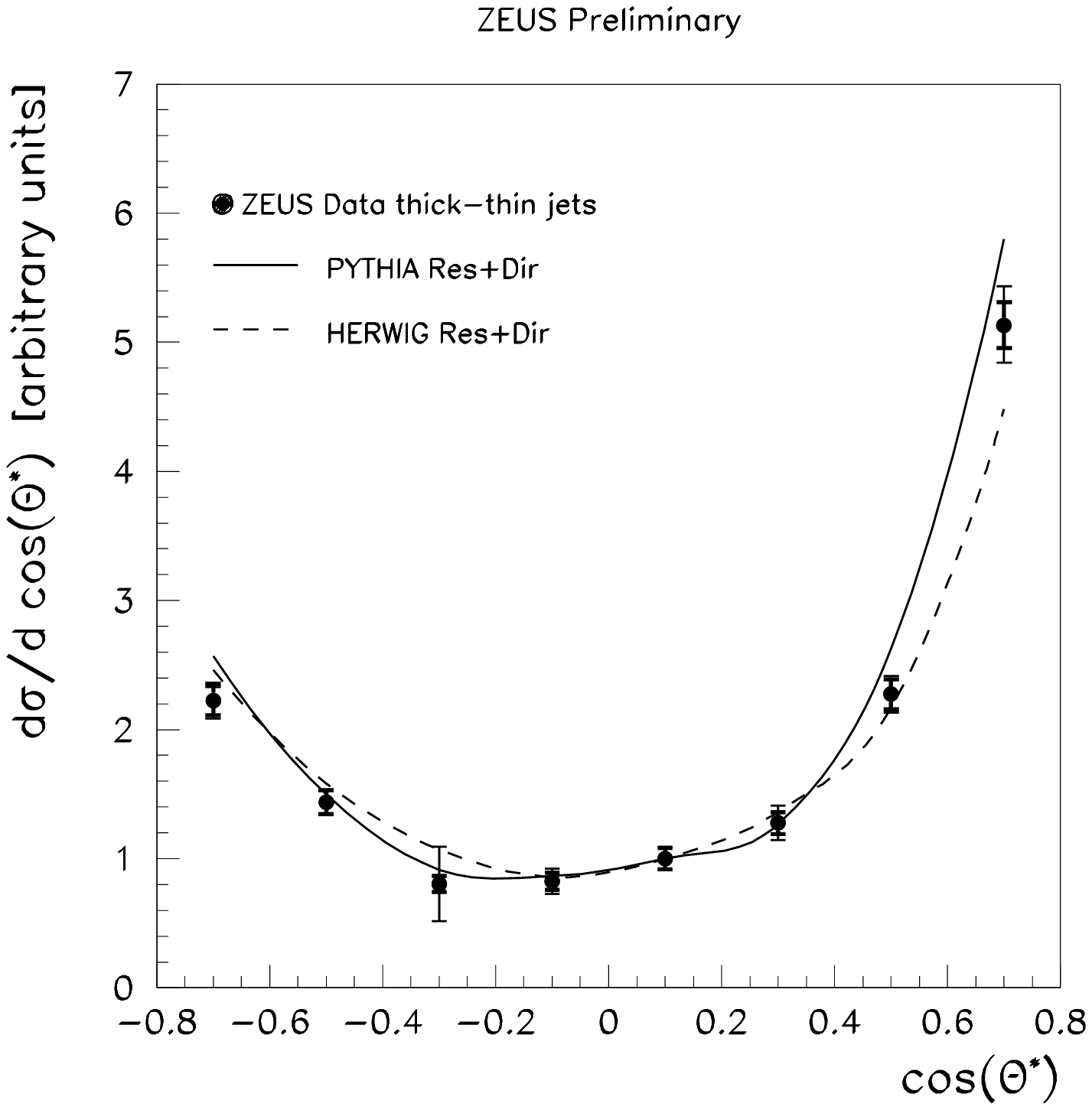,width=10cm}}
\put (5.25,1.24){\epsfig{figure=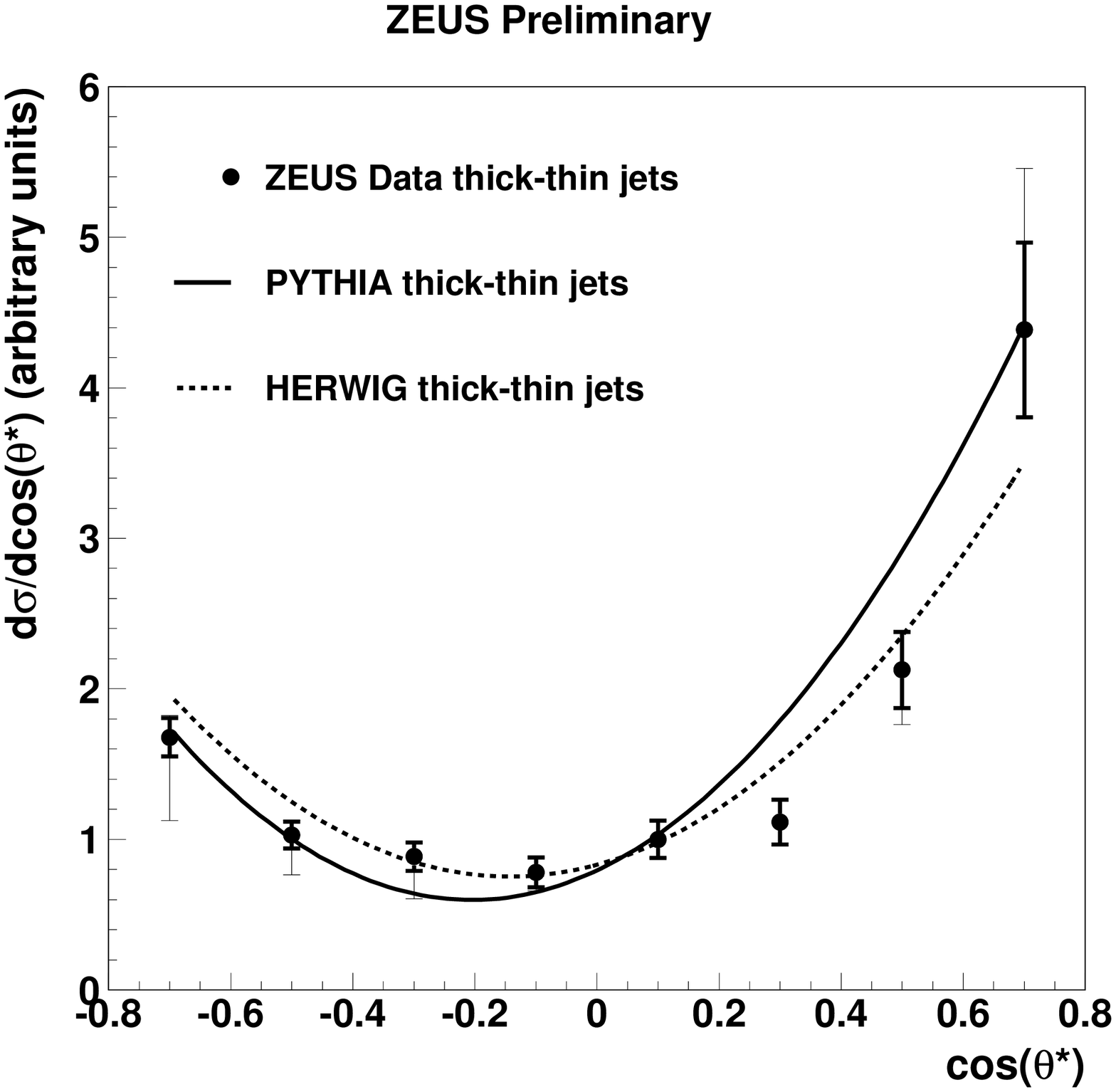,width=6.92cm,height=7.51cm}}
\put (5.1,1.0){\epsfig{figure=\figdir white.eps,width=1.1cm,height=8cm}}
\put (5.5,1.0){\epsfig{figure=\figdir white.eps,width=7cm,height=1.1cm}}
\put (6.3,4.7){\epsfig{figure=\figdir white.eps,width=3.3cm,height=3cm}}
\put (7.0,8.2){\epsfig{figure=\figdir white.eps,width=3.5cm,height=0.5cm}}
\put (4.0,0.0){\epsfig{figure=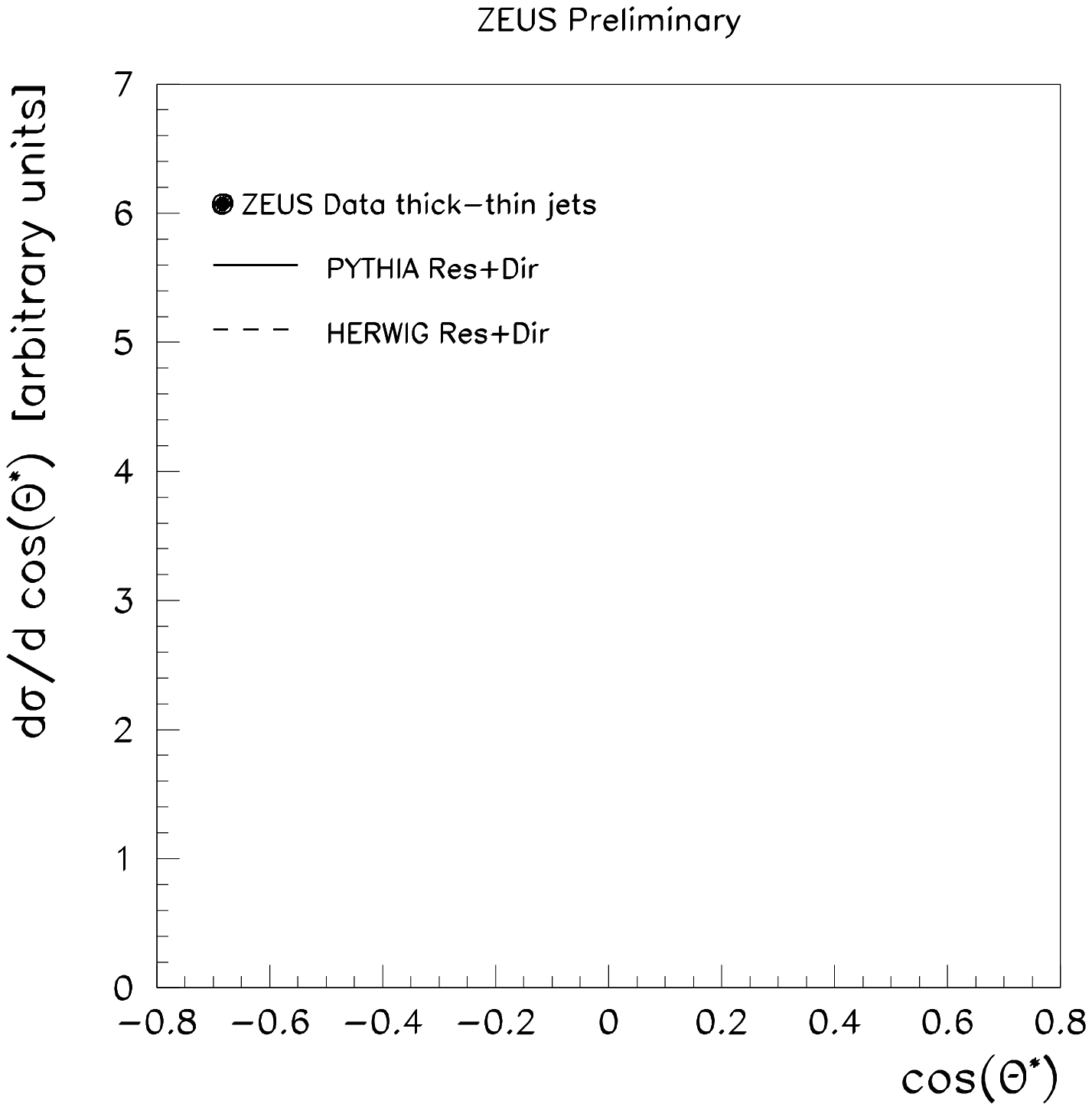,width=10cm}}
\put (-1.7,7.5){\small (a)}
\put (6.4,7.5){\small (b)}
\end{picture}
\end{center}
\vspace{-1.0cm}
\caption{\label{fig7} Measured $\sccos$ for one ``thick'' jet and one
``thin'' jet in an event (black dots) selected according to their shape (a) or
to their subjet multiplicity (b). PYTHIA and HERWIG Monte Carlo calculations
are shown for comparison.}
\end{figure*}

The sample of events with one ``thick'' jet and one ``thin'' jet allows a
measurement of the unfolded $\sccos$ cross section, since in this case the jets
can be distinguished. Figure~\ref{fig7} shows the measured dijet cross section
as a function of $\ccos$. The dijet angular distribution was measured with
respect to the ``thick'' jet and shows a different behaviour on the negative
and positive sides: the measured cross section at $\ccos=0.7$ is more than two
times larger than the one at $\ccos=-0.7$. The Monte Carlo models give a
reasonable description of the shape of the measured $\sccos$
(see figure~\ref{fig7}). The measurements and the Monte Carlo calculations in
figure~\ref{fig7} are normalised so as to have a value of unity at central
values of $\ccos$ to study the difference in slope as $\ccos\rightarrow\pm 1$.
The observed asymmetry is understood in terms of the dominant subprocess:
$q_{\gamma}g_p\rightarrow qg$. The positive side is dominated by $t-$channel
gluon exchange, whereas the negative side is dominated by $u-$channel quark
exchange.

%Figure 7

\section*{Summary and conclusions}

Measurements of jet shape and subjet multiplicity have been performed for
inclusive jet samples with $\etjet>15$ GeV and $-1<\etajet<2$ in the kinematic
regime given by $0.2<y<0.85$ and $\q2\leq 1$~\g2. The measured jet shape
broadens and the mean subjet multiplicity increases as $\etajet$ increases.
Leading-logarithm parton-shower Monte Carlo models with initial and final
state QCD radiation give a good description of the data. The observed
broadening of the jet shape and the increase in the mean subjet multiplicity
as $\etajet$ increases is consistent with an increase of the fraction of gluon
jets.

The Monte Carlo models reproduce the measurements of the jet shape and subjet
multiplicity and display the expected differences for quark- and
gluon-initiated jets and allow the use of the jet shape and subjet
multiplicity to select samples enriched in quark- and gluon-initiated jets to
study the dynamics of the subprocesses. Measurements of the dijet cross
section as a function of $\etajet$ for samples of ``thick'' and ``thin'' jets
show the expected behaviour for samples enriched in gluon- and quark-initiated
jets. The $\cost$ distribution for a two ``thick''-jet sample displays a
similar behaviour to the one expected for a sample enriched in processes
mediated by gluon exchange, whereas that for a two ``thin''-jet sample shows a
behaviour similar to a sample enriched in processes mediated by quark
exchange. Finally, the $\ccos$ distribution for a sample of events with one
``thick'' and one ``thin'' jet exhibits a large asymmetry consistent with the
expected dominance of $t-$channel gluon ($u-$channel quark) exchange as
$\ccos\rightarrow +1(-1)$.

\vspace{0.5cm}
{\bf Acknowledgements.} I would like to thank my colleagues from ZEUS for
their help in preparing this report and the organisers of the conference for
providing a warm atmosphere and hospitality.

\end{document}

%% file: definitions.tex
\def\etjet{E_T^{jet}}
\def\etajet{\eta^{jet}}

\def\etaphi{\eta-\varphi}

\def\qg2{$\q2>125$~\g2}

\def\seta{d\sigma/d\etajet}

\def\q2{Q^2}
\def\pb1{pb$^{-1}$}
\def\gp{\gamma p}

\def\g2{GeV$^2$}

\def\f2g{F_2^{\gamma}}
\def\f2gv{F_2^{\gamma^*}}

\def\rr1{R=1.0}
\def\r7{R=0.7}
\def\R71{R=0.7\ {\rm and}\ 1.0}

\def\mj{M^{JJ}}
\def\m3j{M^{3J}}

\def\cost{\vert\cos\theta^*\vert}

\def\scost{d\sigma/d\cost}
\def\sccos{d\sigma/d\ccos}

\def\kt{k_T}

\def\lq2{\log_{10}(\q2)}

\def\Journal#1#2#3#4{{#1} {#2} (#3) #4}

\def\NPB{{\em Nucl. Phys.} {\bf B}}
\def\PLB{{\em Phys. Lett.}  {\bf B}}

\def\PRD{{\em Phys. Rev.} {\bf D}}

\def\CPC{{\em Comp. Phys. Comm.}}

\def\colab#1{{#1 Collaboration}}

\def\conf#1#2#3#4{{paper #1}, {submitted to the #2}, {#3} {(#4)}}

\def\talk#1#2#3#4#5{{Talk given at the {\it #1}}, {#2}, {#3}, {#4}, {#5}.}

\def\etal{{\it et al}}

\def\ccos{\cos\theta^*}

\def\ro{\rho(r)}
\def\ps{\psi(r)}

\def\ns{n_{subjet}}

\def\yc{y_{cut}}

\def\z0{Z^{\circ}}

\def\oalphas2{O(\alpha\alpha_S^2)}

\def\p2{P^2}

%% file: substructure.bbl
\vspace{-0.25cm}
\begin{references}
\vspace{-0.25cm}

\bibitem{shape}
\colab{ZEUS},
``Measurements of Jet Substructure in Photoproduction at HERA'',
\conf{530}{International Europhysics Conference on HEP}{Tampere}{1999}.

\bibitem{substr}
\colab{ZEUS},
``Substructure Dependence of Dijet Cross Sections in Photoproduction at HERA'',
\conf{424}{XXXth International Conference on HEP}{Osaka}{2000}.

\bibitem{kt}
S. Catani \etal, \Journal{\NPB}{406}{1993}{187};
S.D. Ellis and D.E. Soper, \Journal{\PRD}{48}{1993}{3160};
M.H. Seymour, \Journal{\NPB}{513}{1998}{269}.

\bibitem{pythia}
H.-U. Bengtsson and T. Sj\"ostrand, \Journal{\CPC}{46}{1987}{43};
T. Sj\"ostrand, \Journal{\CPC}{82}{1994}{74}.

\bibitem{herwig}
G. Marchesini \etal, \Journal{\CPC}{67}{1992}{465}.

\bibitem{angular}
\colab{ZEUS}, M. Derrick \etal , \Journal{\PLB}{384}{1996}{401}.
\end{references}
